\documentclass[12pt]{iopart}

\usepackage[utf8]{inputenc}
\usepackage[english]{babel}
\usepackage[T1]{fontenc}
\usepackage{bm}
\usepackage{iopams}
\usepackage{braket}
\usepackage[numbers]{natbib}
\usepackage{tikz}
\usepackage{lipsum}
\usepackage{centernot}
\usepackage{comment}

\begin{document}

\title{Mimicking states with limited resources: \newline passing quantum quiz via global control}

\author{P. V. Pyshkin}
\ead{pavel.pyshkin@gmail.com (corresponding author)}
\address{Physikalisches Institut, Experimentelle Physik III, Universit\"at W\"urzburg, Am Hubland, 97074 W\"urzburg, Germany}
\address{Institute for Topological Insulators, Am Hubland, 97074 W\"urzburg, Germany}
\address{Department of Physical Chemistry, University of the Basque Country UPV/EHU, 48080 Bilbao, Spain}

%\orcid{0000-0003-0017-8880}
\author{E. Ya. Sherman}
\ead{evgeny.sherman@ehu.eus}
%\orcid{0000-0002-2110-2765}
\address{Department of Physical Chemistry, University of the Basque Country UPV/EHU, 48080 Bilbao, Spain}
\address{EHU Quantum Center, University of the Basque Country UPV/EHU}
\address{Ikerbasque, Basque Foundation for Science, 48011 Bilbao, Spain}

\author{A. G\'abris}
\ead{gabris.aurel@fjfi.cvut.cz }
%\orcid{0000-0002-2671-6328}
\address{Czech Technical University in Prague, Faculty of Nuclear Sciences and Physical Engineering B\v rehov\'a 7, 115 19 Praha 1, Star\'e M\v esto, Czech Republic.}
\address{Institute for Solid State Physics and Optics, Wigner Research Centre, P.O. Box 49, H-1525 Budapest, Hungary}

\author{Lian-Ao Wu}
\ead{lianaowu@gmail.com}
\address{Department of Physics, University of the Basque Country UPV/EHU, 48080 Bilbao, Spain}
\address{EHU Quantum Center, University of the Basque Country UPV/EHU}
\address{Ikerbasque, Basque Foundation for Science, 48011 Bilbao, Spain}

%\orcid{0000-0003-4896-6958}

%\maketitle

\vspace{10pt}
\begin{indented}
	\item[]October 2023
\end{indented}

\begin{abstract}
  Precise control of quantum systems with a moderate number of degrees of freedom,
  being of interest for application in quantum technologies, becomes experimentally feasible. 
  Various types of quantum scenarios and protocols are being widely discussed in scientific 
  literature. We propose, analyze, and optimize a protocol which allows fast simulation of 
  properties of unknown quantum states relying on minimum relevant information. 
  Our protocol, having common features with quantum identification and shortcuts to adiabaticity,
  permits avoiding orthogonality catastrophe, where transitions between physically very 
  similar systems are characterized by zero or a very low fidelity.    
\end{abstract}

\section{Introduction: quantum identification and state similarity}

The problem of quantum identification~\cite{q_identif_1, discrimination_Bae_2015,  Identification-Pyshkin2018, State-discrimination-PhysRevA.95.023828} 
can be described as follows. Device that can produce a quantum system in 
a pure state~$\ket{\psi_{l}}$ belonging 
to a known set of possible states $\{ \ket{\psi_{1}}, \ket{\psi_{2}}, \dots, \ket{\psi_L}  \}$, 
where $L$ is the total number of those states, i.e. $1\leq l\leq L$. 
The problem is to identify $l$ by using the minimal possible resources.   
This large task can be modified as follows: instead of identifying $l$ one can 
{\em prepare} a state with a feature similar to one of the features characterizing the $\ket{\psi_{l}}$ state. 
Such a preparation has to use limited resources: 
time and information about $\ket{\psi_{l}}$ and/or a restricted set of permitted quantum operations. The preparation of desired quantum state means that we apply some permitted quantum operations to quantum system being in some ``trial'' state. Therefore this procedure can be viewed as a quantum control problem of a special kind~\cite{Koch2022, QuantControl, Quantum-Control-numerical-book}. Because time is a limited resource we notice 
that our protocol may be related to shortcuts to adiabaticity~\cite{Shortcuts-Torrontegui-Muga-2013}  and quantum speed limit problem~\cite{Pyshkin-PRA-2019} .
%means finding a protocol of evolution of quantum system from its initial state $\ket{\psi_i}$ to some demanded final state $\ket{\psi_f} = U(T)\ket{\psi_i}$ with finite (short) time~$T$. It is known that there is transformation $U^{\prime}$, which generated by adiabatically slow process, where $\ket{\psi_f} = U^{\prime}\ket{\psi_i}$.

Usually, in order to determine success metric of a quantum protocol one uses the fidelity 
$|\braket{\psi|\psi^{\prime}}|^2$, where $\ket{\psi^{\prime}}$ is a desired state and $\ket{\psi}$ is a state which is the
result of the protocol. 
However, such a strict condition of success can be excessive for some applications and may require an exponentially long search time. 
Therefore we consider the following scenario in this paper.
We have two complex systems, for example, two spin chains with the same number of spins~$N$. 
The state of the first (second) chain is $\ket{\psi_{1}}$ ($\ket{\psi_{2}}$), and their fidelity~$|\braket{\psi_{1}|\psi_{2}}|^2 = 1$.
Let us extend both chains by adding to them one spin in a state $\ket{\phi_{1}}$ ($\ket{\phi_{2}}$) for the 
first (second) chain, thus, we have two chains with $N+1$ spins being in 
the states~$\ket{\Psi_{1}} = \ket{\psi_{1}}\otimes\ket{\phi_{1}}$ ($\ket{\Psi_{2}} = \ket{\psi_{2}}\otimes\ket{\phi_{2}}$) for first (second) chain.
In this scenario the common fidelity $|\braket{\Psi_{1}|\Psi_{2}}|^2 = |\braket{\phi_{1}|\phi_{2}}|^2$ is 
defined by the fidelity of these additional spins only.
If $\braket{\phi_{1}|\phi_{2}} = 0$ the state of the first $N$ spins in each chain becomes irrelevant.
This is similar to the well-known 
orthogonality catastrophe~\cite{ort-catastr-PhysRevLett.18.1049, ort-catastr-PhysRevLett.124.110601}.

However, we notice that in quantum control procedures, especially in the tasks of quantum state preparation 
it is important to operate with such a measure of closeness of quantum states 
which takes into account similarities 
of different parts of the system (alongside with the entanglement measure similar to that proposed in~\cite{Scott2004}).  
In this report we introduce and propose to apply a less rigorous than the conventional fidelity
metric to a complex system which consists of distinguishable parts. 
Application of such a metric can be considered as a ``quantum quiz'' consisting of 
several ``questions'' with answers to each of them being related to the similarity of the corresponding parts 
of the system. Optimization of this metric is used as an intermediate feedback as well as the final goal.

\section{The setting}

\begin{figure}[t]
	\centering
	\includegraphics[width=80mm]{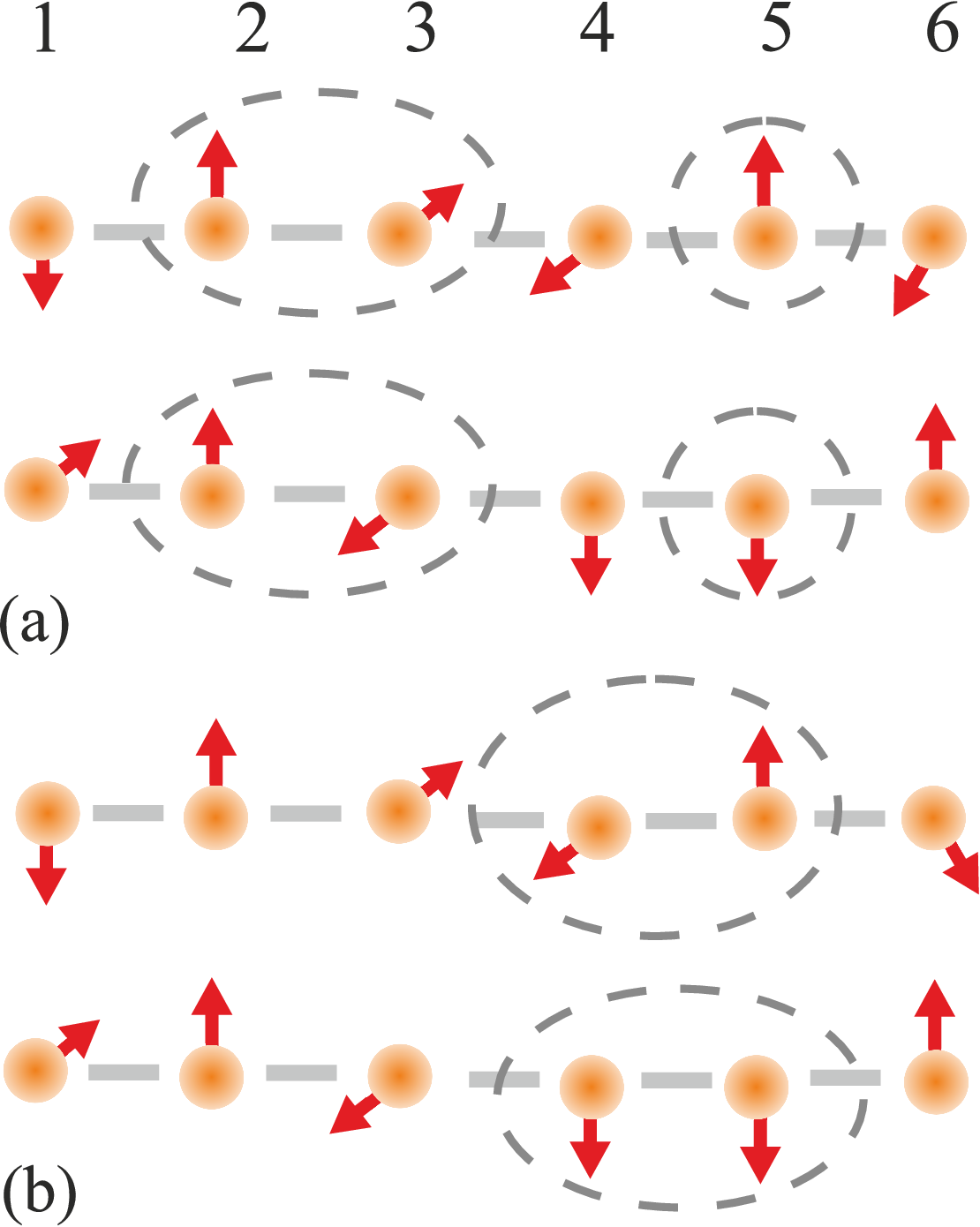}
	\caption{Example of choosing subsystems. Here have $N=6$ spins in both target and candidate systems, 
	and different values of~$\{\bm k\}$ in~(\ref{general_fidelity}) correspond
	to sets of spins including $\{\bm k\} = \{ 2, 3, 5\}$ (with $\overline{\{\bm k\}} = \{1,4\}$) 
	(panel a), and set of spins $\{\bm k\} = \{4,5\}$ (with $\overline{\{\bm k\}} = \{1,2,3\}$) (panel b).}
	\label{fig:similarity}
\end{figure}

In this section we introduce the problem in general terms.
Let us assume we have a complex quantum system which consists of $N$ {\em distinguishable} 
{\em enumerated} parts, and this system is 
in its ground state~$\ket{\psi_0}$ corresponding to Hamiltonian $H = H(b_{1}, b_{2}, \dots, b_{K})$, where $b_{i}$ are unknown 
parameters, and $K$ is the number of these parameters. 
This system we call as {\em target system}.
Let us now introduce the second system with the same nature, i.e. a system consisting of~$N$ ($N\geq2$) 
non-divisional parts with the same 
dimension of the Hilbert space for all parts as the target system. For example, 
if the target system is $N$ spins-$1/2$, the second system is $N$ spins-$1/2$ (or $N$ qubits) too. 
Here we call the second system a {\em candidate} system. The candidate system initially is supposed 
to be in the ground state~$\ket{\psi^{\prime}_0}$ of the Hamiltonian~$H=H(b^{\prime}_{1}, b^{\prime}_{2}, \dots, b^{\prime}_{K})$. 
In order to compare these two states we introduce the {\em similarity} measure
\begin{equation}\label{general_fidelity}
	F(\ket{\psi^{[t]}}, \ket{\psi^{[c]}}) = \sum_{\{\bm k\}} f_{\{\bm k\}}({\bm\rho}_{\{\bm k\}}^{[t]}, {\bm\rho}_{\{\bm k\}}^{[c]}),
\end{equation} 
where $\{\bm k\}$ labels a set of subsystems (each set can contain from $1$ to $N-1$ subsystems, see Fig. \ref{fig:similarity}), 
the maximal number of terms in sum is $2S(N,2)$, where $S(N,2)=2^{N-1}-1$ is the number of all the possible bipartitions 
of $N$ subsystems defined via Stirling number of the second kind, ${\bm\rho}_{\{\bm k\}}^{[t]}$ (${\bm\rho}_{\{\bm k\}}^{[c]}$) 
is the reduced density matrix of subsystem $\{\bm k\}$ of the target (candidate) 
system, and $f_{\{\bm k\}}$ are functions which can be different for different subsystems. 
%If Fig. \ref{fig:similarity} we show two examples of choosing subsystem sets in~(\ref{general_fidelity}): we have $N=6$ spins in both target and candidate systems, and different values of~$k$ in~(\ref{general_fidelity}) corresponds to sets of spins including $\{ 2, 3, 5\}$ (panel a), and set of spins $\{4,5\}$ (panel b).  

The reduced density matrices of subsystems~$\{\bm k\}$ are defined as partial 
traces over all other subsystems except the~$\{\bm k\}$th set :
\begin{equation}\label{rho_{K}}
	{\bm\rho}_{\{\bm k\}}^{[t,c]} = \tr_{\overline{\{\bm k\}}} \ket{\psi^{[t,c]}}\bra{\psi^{[t,c]}},
\end{equation} 
see Fig. ~\ref{fig:similarity}. We assume that the similarity~$F$ reaches 
the maximum at ${\bm\rho}_{\{\bm k\}}^{[t]} = {\bm\rho}_{\{\bm k\}}^{[c]}, \forall \{\bm k\}$, 
although this is not a necessary condition to achieve the maximum. 

Next we assume that there is a limited resource- ``blackbox'' device  which can compare the candidate and the target systems, and 
output values of the similarity~(\ref{general_fidelity}). 
The task is to implement some unitary operation~$U$ to candidate system in order to 
have $F(\ket{\psi_0^{[t]}}, U\ket{\psi_0^{[c]}})>F(\ket{\psi_0^{[t]}}, \ket{\psi_0^{[c]}})$ with choosing~$U$ 
in the most optimal way to have
\begin{equation}\label{general_task}
	F(\ket{\psi_0}^{[t]}, U\ket{\psi_0^{[c]}}) = \max_{U^{\prime}}F(\ket{\psi_0^{[t]}}, U^{\prime}\ket{\psi_0^{[c]}}),
\end{equation} 
where $U^{\prime}$ belongs to a family of allowed for implementation unitary transformations. We assume that $U = U_{M}U_{M-1}\dots U_{1}$. 
This means, that it is possible to apply $M$ different unitary operations, and after applying each of them we 
use the blackbox in order to estimate the current value of similarity~(\ref{general_fidelity}) which can be used to choose 
the next unitary operation. We assume all these unitary operators  $U_{j}$ ($j=1,2,\dots M$) are generated by special-types of Hamiltonians with fixed or limited evolution time or energy. 
In analogy with ``quantum quiz'', we allow to
reexaminate~$M$ times, and our goal is to increase similarity~(\ref{general_fidelity}) as much as possible 
with $M$ being as less as possible (see Introduction). 

The definition~(\ref{general_fidelity}) can be considered as the final
grade of ``quantum quiz'' consisting of~$2S(N,2)$ ``questions'', where each ``question'' is
related to a certain part of complex quantum system. 
Similar kind of measure which involves density matrices of parts of quantum system has been presented
earlier in Ref. \cite{Scott2004} to analyze entanglement properties. 
In the next section we present and consider in detail a realization of the problem described above.

\section{Mimicking states in a spin chain}

\subsection{Analytical procedure and numerical analysis for discrete distribution}

As an example we consider periodic spin-1/2 chain of~$N$ spins with the following Hamiltonian

\begin{equation} \label{Hchain}
	H = \sum_{k=1}^{N}(X_{k} + b_{k}Y_{k} + JZ_{k}Z_{k+1}), 
\end{equation}
where $X,Y,Z$ are Pauli matrices and the first and last sites to be adjacent with $N+1\equiv1$. Further we assume antiferromagnetic ~$J=1$.  

\begin{figure}
	\centering
	\includegraphics[width=80mm]{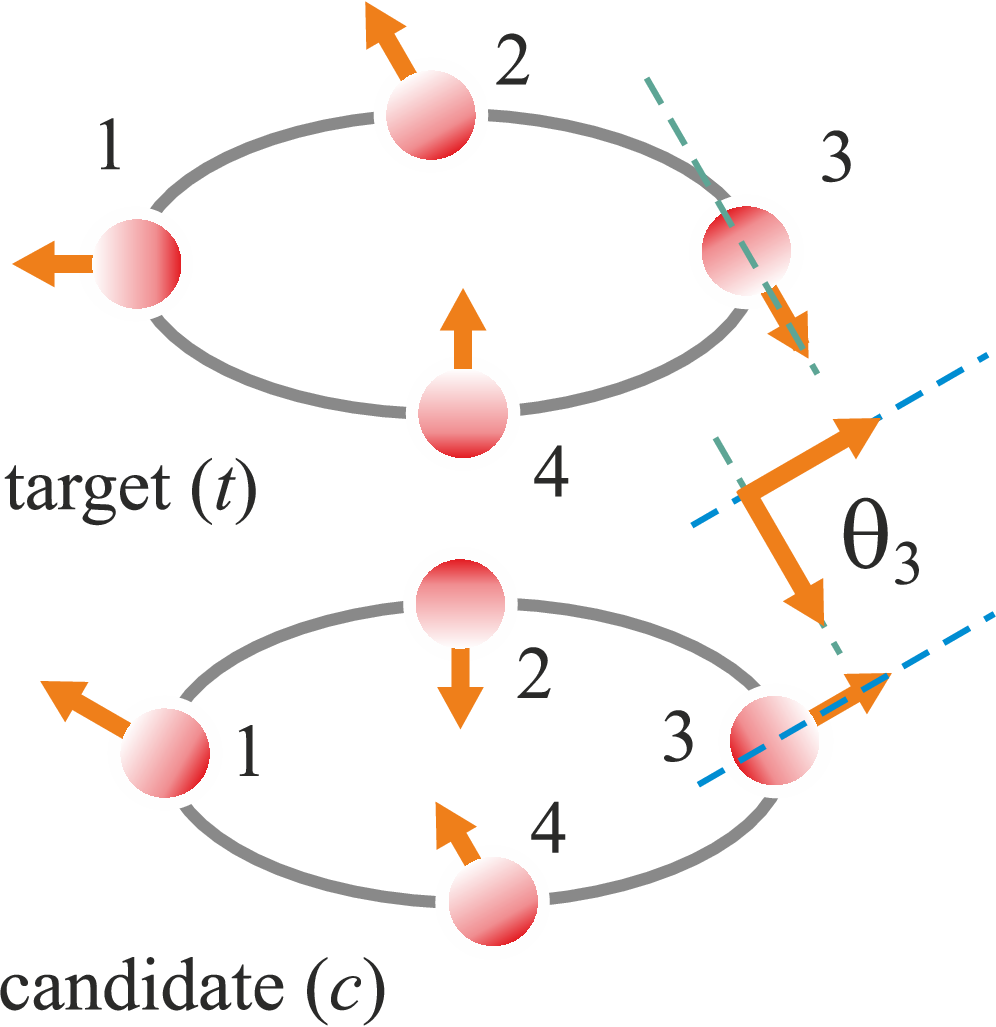}
	\caption{$N=4$ periodic spin chains as a target (upper panel) and candidate (lower panel) systems.}
	\label{fig:figure1}
\end{figure}

In this example we will only compare each spin in the target system with the corresponding one in the candidate system, 
thus we can mark~$\{\bm k\} = k$, where~$1\leq k\leq N$ in~(\ref{general_fidelity}), %and $k$ is a number of spin in the chain
\begin{equation}\label{fidelity_for_chain}
	F(\ket{\psi^{[t]}}, \ket{\psi^{[c]}}) = \sum_{k=1}^{N} \cos\theta_{k},
\end{equation}  
where $\theta_{k}$ is the angle between the Bloch vectors of the $k$th target 
spin and $k$th candidate spin~(see Fig. \ref{fig:figure1}). 
Below we show that~$\cos\theta_{k}$ is a function of ${\bm\rho}_{k}^{[t]}$ and ${\bm\rho}_{k}^{[c]}$ 
as in definition~(\ref{general_fidelity}).  
 
The density matrices~(\ref{rho_{K}}) of spins are expressed in terms of their Bloch vectors
\begin{equation}\label{Bloch_vecs}
	\theta_{k} = \widehat{\bm{r}_{k} \bm{c}_{k}}, \quad {\bm\rho}^{[t]}_{k} = 
	\frac{\mathbb{I} + \bm{r}_{k}\bm{\sigma}_{k}}{2}, \quad {\bm\rho}^{[c]}_{k} = 
	\frac{\mathbb{I} + \bm{c}_{k}\bm{\sigma}_{k}}{2},
\end{equation}
where $\bm{r}_{k}$ ($\bm{c}_{k}$) is the Bloch vector of the $k$th spin of the target 
(candidate) system, and $\bm{\sigma}_{k} = (X_{k}, Y_{k}, Z_{k})$ is the 
Pauli vector for the $k$th spin, $\mathbb{I}$ is identity matrix.

We now consider a family of the allowed unitary operations in~(\ref{general_task}) 
determined by $U = \exp(-iH_{1}t)$ generated by the Hamiltonian
\begin{equation}\label{Hrotation}
	H_{1} = \bm{B} \cdot  \sum_{k=1}^{N}\bm{\sigma}_{k},
\end{equation}
where $\bm{B}$ is a vector parameter the same for all spins~\cite{Global-field-gate-PhysRevLett.93.030501}. 
The Hamiltonian~(\ref{Hrotation}) is not assumed to be able to provide the {\em complete} 
control~\cite{controllability-PhysRevA.79.060305}, therefore we do not expect 
the maximal possible similarity as a result of our algorithm. 
We assume $\bm{B} = B \bm{b}$, with $B\geq 0$ and $|\bm{b}| = 1$.

Notice that for our setting since the Bloch vectors do not have $z$-component, 
this allows us to use $\bm{b} = (0,0,\pm1)$ in order to have spin rotations 
generated by~(\ref{Hrotation}) only in the $XY$ plane. Here we present main algebraic relations 
which will be used to understand the optimization and the numerical simulation procedures in the next subsection.

In order to show that our choice of similarity~(\ref{fidelity_for_chain}) does 
not contradict the definition~(\ref{general_fidelity}) we show 
that the cosine function of the angle $\theta_{k}$ ($\cos\theta_{k} = {\bm{r}_{k}\bm{c}_{k}}/{{r}_{k}{c}_{k}}$) between average spins number~$n$ of the target and candidate systems can be 
expressed in the following way. 
%From the definition of the scalar product we have
%\begin{equation}\label{cos-2}
%	\cos\theta_{k} = \frac{\bm{r}_{k}\bm{c}_{k}}{{r}_{k}{c}_{k}}.
%\end{equation} 
With using~(\ref{Bloch_vecs}) and algebra of the Pauli matrices we write
\begin{equation}\label{tr_rhorho}
	\tr {\bm\rho}^{[t]}_{k}{\bm\rho}^{[c]}_{k} = \frac{1 + \bm{r}_{k}\bm{c}_{k}}{2}.
\end{equation} 
Then from~(\ref{tr_rhorho}) we have
\begin{equation}\label{cos-1}
\cos\theta_{k} = \frac{2\tr({\bm\rho}^{[t]}_{k}{\bm\rho}^{[c]}_{k})-1}
{\sqrt{2\tr\left({\bm\rho}^{[c]}_{k}\right)^2 - 1}
\sqrt{2\tr{\left({\bm\rho}^{[t]}_{k}\right)}^{2} - 1}},
\end{equation} 
where all rotations generated by~(\ref{Hrotation}) do not affect the denominator in~(\ref{cos-1}).

For a global rotation governed by~(\ref{Hrotation}) we introduce $\chi = Bt$ and obtain $V = \exp(-iH_{1}t)$ $=\prod_{k=1}^N V_{k}$ 
$=\prod_{k=1}^N\exp(-i\bm{b}\bm{\sigma}_{k}\chi)$, thus
$$
\tr_{\overline{k}} (V\ket{\psi}\bra{\psi}V^\dagger) = V_{k}{\bm\rho}_{k}V_{k}^\dagger.
$$ 
This allows us to express the changes of similarity after the rotation applied to 
the candidate system in terms of the Bloch vectors before the rotation and parameters ${\bm b}$,~$t$~and~$B$
\begin{comment}
\begin{widetext}
\begin{multline}
	F(\ket{\psi^{[t]}}, V\ket{\psi^{[c]}})-F(\ket{\psi^{[t]}}, \ket{\psi^{[c]}}) = \\
	2\sin t \sum_{k=1}^{N}\left[\sin t \left(\vphantom{\frac{1}{1}} (\frac{\bm{c}_{k}}{{c}_{k}}\bm{B})( \frac{\bm{r}_{k}}{{r}_{k}}\bm{B}) - \frac{\bm{c}_{k}}{{c}_{k}}\frac{\bm{r}_{k}}{{r}_{k}} \right) + \bm{B}( \frac{\bm{c}_{k}}{{c}_{k}}\times\frac{\bm{r}_{k}}{{r}_{k}} ) \cos t \right],
\end{multline}
\end{widetext}
\end{comment}

%\begin{widetext}
	\begin{eqnarray}\label{DeltaF}
		F(\ket{\psi^{[t]}}, V\ket{\psi^{[c]}})-F(\ket{\psi^{[t]}}, \ket{\psi^{[c]}}) = \nonumber \\
		\quad 2\sin \chi \sum_{k=1}^{N} \sqrt{\left( 1 - \left(\frac{\bm{c}_{k}}{c_{k}}\bm{b}\right)^2 \right)\left( 1 - \left(\frac{\bm{r}_{k}}{r_{k}}\bm{b}\right)^2 \right)} \sin (\chi + \gamma_{k}),
	\end{eqnarray}
%\end{widetext}
with
\begin{equation}
	\gamma_{k} = \arccos \frac{(\bm{c}_{k}\bm{b})(\bm{r}_{k}\bm{b}) - \bm{c}_{k}\bm{r}_{k}}{\sqrt{\left( c_k^2 - \left({\bm{c}_{k}}\bm{b}\right)^2 \right)\left( r_k^2 - \left({\bm{r}_{k}}\bm{b}\right)^2 \right)}}  
\end{equation}
where we use~(\ref{cos-1}) and~$\exp(-i\bm{b}\bm{\sigma}_{k}Bt) = \mathbb{I}\cos (Bt) - i (\bm{b}\bm{\sigma}_{k})\sin (Bt)$. 
Since in our system~$\bm{b}\perp\bm{c}_{k},\bm{r}_{k}$, the square root in~(\ref{DeltaF}) 
is equal to~$1$ and $\gamma_{k} = \arccos(-(\bm{c}_{k}\bm{r}_{k})/(c_{k}r_{k})) = \pi - \theta_{k}$. 
As result~(\ref{DeltaF}) can be rewritten as 
\begin{equation}\label{DeltaF2}
	F(\ket{\psi^{[t]}}, V\ket{\psi^{[c]}})-F(\ket{\psi^{[t]}}, \ket{\psi^{[c]}}) = 
	 2\sin \chi \sum_{k=1}^{N} \sin (\theta_{k} - \chi).
\end{equation}
Notice that if $\theta_{k}=\theta$, then~(\ref{DeltaF2}) gives 
the condition for the maximal increase in the similarity: $\chi = \theta/2$, meaning  
that we have to consider positive angles $\theta_{k}$ for the case when 
vector $\bm{c}_{k}$ must be rotated counter clock-wise in order to decrease~$\theta_{k}$.
%Our fidelity measure~(\ref{general_fidelity}) contains sum of cosines 

For simplification, we further assume that possible values of parameters in~(\ref{Hchain}) are 
discrete (each $b_{k}$ can take only $D$ different values) and relatively small, 
for example, $|b_{k}|\leq1$, and thus we expect $\theta_{k}<2\arctan1=\pi/2.$ 
%and this is why it is convenient to consider sum of Bloch vectors of all spins in chains: 
%$\bm{C} = \sum_{k}^N \bm{c}_{k}$ and $\bm{R} = \sum_{k}^N \bm{r}_{k}$.
%Note, the lengths of~$\bm{c}_{k}$ and~$\bm{r}_{k}$ depend on strength of~$ZZ$ coupling in~(\ref{Hchain}), while their directions and angles~$\theta_{K}$ between them don^{\prime}t. Thus here we compare states of possibly highly entangled systems totally ignoring entanglement properties in our fidelity metric.   
%It is obvious $F=N \implies \bm{R}\parallel\bm{C}$ (however, $\bm{R}\parallel\bm{C} \centernot\implies F=N$). Hamiltonian~(\ref{Hrotation}) produces global rotation of the candidate vector~$\bm{C}$, and our goal now is to organize rotation of~$\bm{C}$ to its direction coincides with~$\bm{R}$. The problem is to determine the right angle of rotation (in other words, determine evolution time~$t$ for applying Hamiltonian~(\ref{Hrotation})) with using as less information on fidelity~(\ref{general_fidelity}) as possible. 

\begin{figure}[t]
	\centering
	\includegraphics{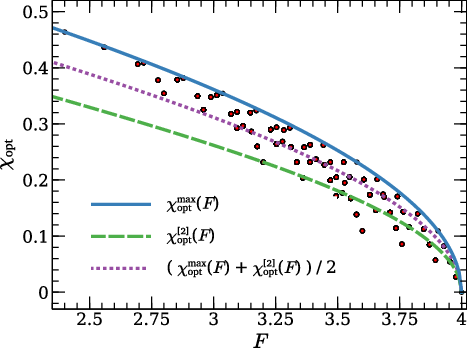}
	\caption{Optimal rotation angle as a function of similarity from the blackbox query for $N=4$ and $b_k$ being in set $\{-0.5,-0.25,0,0.25,0.5\}$. Circles correspond to each possible combination of parameters~$\bm{b}$, lines represent averaged values of optimal angle given by Eqs.(\ref{chi_opt_1}) and (\ref{chi_opt_2}).}
	\label{fig:figure2}
\end{figure}

Before discussing the algorithm we produce a special statistics for all possible target states with an initial state~$\ket{\psi_0^{[c]}}$ corresponding to $b_k = b_{\min}$, where $b_{\min}$ 
is the minimal possible value of parameters~$b_{k}$. 
This assumption leads to all angles~$\theta_{k}\geq0$.
This numerics can be done without using resources of blackbox. 
For each possible target state~$\psi_0^{[t]}$ we calculate angle~$\chi_{\rm opt}$ which corresponds to the optimal global rotation
\begin{equation}\label{U-rotation}
	U = \exp\left(-i \chi_{\rm opt} \sum_{k} Z_{k}\right),
\end{equation}
which maximizes~$F(\ket{\psi_0}^{[t]}, U\ket{\psi_0^{[c]}})$. The optimal angle is given by the {\em circular mean}
\begin{equation}\label{chi-def}
	\chi_{\rm opt} = \frac{1}{2}\arctan\left( \frac{1}{F}\sum_{k=1}^N\sin\theta_{k}  \right),
\end{equation}
where $F$ is the similarity from~(\ref{fidelity_for_chain}). 
Expression~(\ref{chi-def}) can be obtained by setting the 
derivative of~(\ref{DeltaF2}) to zero with respect to~$\chi$. 
For large possible values of $|b_{k}|>1$ the similarity~$F$ can become negative, 
and one should use the function ${\mathrm{atan2}}(\sum_{k=1}^N\sin\theta_{k}, F)$ instead of~(\ref{chi-def}).
The result of this statistics is represented as circles in Fig.\ref{fig:figure2}.
This calculated in advance data can be used in the algorithm for optimal similarity increasing. 
However, this approach becomes inefficient for large systems, or when allowed states form continuum. 
Thus we have to find approximate dependence $\chi_{\rm opt} = \chi_{\rm opt}(F)$ for further using it in the algorithm.
In order to do this we modify Eq.(\ref{chi-def}), and express the sum of $\sin\theta_{k}$ through~$F$.
This can be done only for ensemble average. 
Let us assume for a moment that the angles~$\theta_{k}$ are random.
Our goal is to find the following function
\begin{equation}\label{sin-aver}
	\chi_{\rm opt} = \frac{1}{2}\arctan\left( \frac{1}{F} \left\langle \sum_{k=1}^N\sin\theta_{k}  \right\rangle \right).
\end{equation}
Note, due to our choice of initial state the average in angle brackets is not zero when $0\leq \theta_{k} \leq \pi$. 
Also, not all $\theta_{k}$ are independent due to the condition 
\begin{equation}\label{sum_cos}
\sum_{k=1}^N\cos\theta_{k} = F.
\end{equation}
\begin{comment}
Exact calculation of~(\ref{sin-aver}) together with~(\ref{sum_cos}) seems difficult and therefore we try to pick up some ``characteristic'' values from the sets $\sum_{k}\sin\theta_{k}$ for fixed~$F$.
First, in order to find maximal value of $\chi_{\rm opt}$ we solve system of $N+1$ equations $\partial L / \partial \theta_{k} = 0$, $\partial L / \partial \lambda = 0$, with Lagrange function $L = \sum_{k}\sin\theta_{k} - \lambda(\sum_{k}\cos\theta_{k} - F)$ (here we use monotonic growing of $\arctan$ function). We obtain solution $\cos\theta_k = F/N$ for all $k$, and as a result the maximal value of optimal angle is
\begin{equation}\label{chi_opt_1}
	\chi_{\rm opt}^{\rm max} = \frac{1}{2}\arctan\left( \sqrt{\frac{N^2}{F^2} - 1}  \right).
\end{equation}
The curve corresponding to~(\ref{chi_opt_1}) is shown in Fig.\ref{fig:figure2} as solid line.
The expression~(\ref{chi_opt_1}) corresponds to extreme symmetric situation, when all angles~$\theta_{k}$ are the same.
This give us a clue to consider next a quite asymmetric situation for angles~$\theta_{k}$.
Now we assume $N$ is even, and $\theta_{1, 2, \dots, N/2} = 0$, and for $k>N/2$ we assume $\cos\theta_{k} = 2F/N - 1$ (it is easy to check that this choosing of~$\theta_{k}$ satisfies~(\ref{sum_cos}) ). Then we have another dependence for optimal angle picked from the ensemble of possible combinations of~$\theta_{k}$:
\end{comment}

The exact calculation of~(\ref{sin-aver}) in conjunction with~(\ref{sum_cos}) presents a challenge, leading us to extract some "characteristic" values from the set of all possible sums \(\sum_{k}\sin\theta_{k}\) for a fixed \(F\). To determine the maximum value of \(\chi_{\rm opt}\), we solve a system of \(N+1\) equations: \(\partial L / \partial \theta_{k} = 0\) and \(\partial L / \partial \lambda = 0\), with the Lagrange function \(L = \sum_{k}\sin\theta_{k} - \lambda(\sum_{k}\cos\theta_{k} - F)\), leveraging the monotonic increase of the \(\arctan\) function. The solution yields \(\cos\theta_k = F/N\) for every \(k\), resulting in the maximal value of the optimal angle being 
\begin{equation}\label{chi_opt_1}
	\chi_{\rm opt}^{\rm max} = \frac{1}{2}\arctan\left( \sqrt{\frac{N^2}{F^2} - 1} \right).
\end{equation}
Figure~\ref{fig:figure2} displays this relationship as a solid line, representing an extreme symmetric case where all angles \(\theta_{k}\) are the same. This prompts us to explore a highly asymmetric scenario for the angles \(\theta_{k}\).
Assuming \(N\) is even, we set \(\theta_{1, 2, \ldots, N/2} = 0\), and for \(k > N/2\), \(\cos\theta_{k} = 2F/N - 1\), which satisfies Equation~(\ref{sum_cos}). This gives rise to another dependence for optimal angle picked from the ensemble of possible combinations of~$\theta_{k}$:
\begin{equation}\label{chi_opt_2}
	\chi_{\rm opt}^{[2]} = \frac{1}{2}\arctan\left( \sqrt{\frac{N}{F} - 1}  \right).
\end{equation}
The curve corresponding to~(\ref{chi_opt_2}) is shown in Fig.\ref{fig:figure2} as dashed line. 
From Fig.\ref{fig:figure2} we see that the average function $\chi_{\rm opt}^{\rm aver} = (\chi_{\rm opt}^{\rm max}+\chi_{\rm opt}^{[2]})/2$ must be a good choice as a guide for choosing optimal unitary after getting information about value of~$F$. In other words, for every value of~$F$ we change the real average over all possible combinations of~$\theta_{k}$ to the average between two ``special'' combinations of~$\theta_{k}$.

We now propose the following protocol for increasing~$F$ with just a single query~$M=1$ to the blackbox:

1.~Let us choose the initial state~$\ket{\psi_0^{[c]}}$ with all the Bloch vectors 
having direction corresponding to $b_k = b_{\min}\;\forall k$, where $b_{\min}$ 
is the minimal possible value of parameters~$b_{k}$.

2.~Ask blackbox for the similarity~$F$.  

3.~After obtaining the result, we calculate corresponding $\chi = \chi_{\rm opt}^{\rm aver}(F)$ and produce rotation~(\ref{U-rotation}).

\subsection{Numerical analysis for continuous distributions}

We restrict ourself by choosing $N=K=4$ and the values of $b_{k}$ from the interval~$(-0.5, 0.5)$. 
Following the protocol we choose the initial state~$\ket{\psi^{[c]}_0}$ corresponding to $b_k = -0.5$ for all~$k$. 

\begin{figure}[t]
	\centering
	\includegraphics{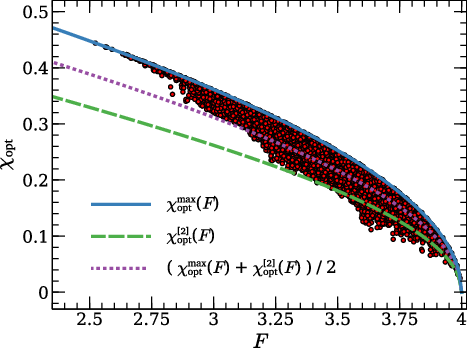}
	\caption{Optimal rotation angle as a function of similarity from the blackbox query for $N=4$ and $b_k$ being random value uniformly distributed in the interval $(-0.5,0.5)$. Circles correspond to optimal angle for each random target state, curves represent angles given by Eqs.(\ref{chi_opt_1}), (\ref{chi_opt_2}) and their mean value.}
	\label{fig:randomsimulation}
\end{figure}

We produced numerical simulation for $10^4$ random target states and for each of these states we 
found optimal rotation angle~$\chi$ as shown in Fig. \ref{fig:randomsimulation}. 
For each target state we calculated the increase in the similarity $\Delta F$ after the 
optimal rotation~(\ref{fig:randomsimulation}).  
Increase in the similarity averaged over all possible target states approximates to~$\langle \Delta F \rangle \approx 0.439$, while with optimal rotational angles we have~$\langle \Delta F \rangle \approx 0.443$.
%It is obvious that for some target states the increase in the similarity is maximally possible, and corresponding points are lying on the line $\Delta F + F = N$.  

%One can decide that for our problem it is possible to choose some optimal angle $\chi_{\rm const}>0$, then rotate any state without asking for similarity value, and similarity will increase. In Fig.\ref{fig:constchi} we show that our algorithm which uses knowledge of similarity is more efficient than using optimal constant parameter~$\chi = \chi_{\rm const}$.

\begin{figure}[t]
	\centering
	\includegraphics{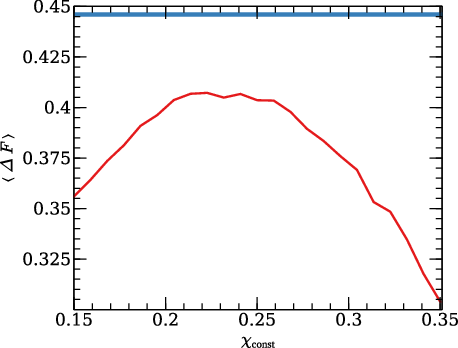}
	\caption{Comparison of average similarity increasing with using our algorithm and rotating initial state by some fixed angle~$\chi_{\rm const}$. The upper line corresponds to the average increase in the similarity by using our algorithm based on~(\ref{chi_opt_1}) and (\ref{chi_opt_2}), while bottom line is the average increase of the similarity by using the same rotation angle~$\chi_{\rm const}$ for any value of similarity~$F$.}
	\label{fig:constchi}
\end{figure}

One might determine that for our problem, selecting some fixed optimal angle~$\chi_{\rm const}>0$ is feasible. By rotating any state with this angle, without the need to evaluate the similarity, an increase in similarity is guaranteed. Figure~\ref{fig:constchi} illustrates that our algorithm, which leverages the knowledge of similarity, is more effective than the approach of employing a constant optimal parameter~$\chi = \chi_{\rm const}$ which gives a lower~$\langle \Delta F \rangle \approx 0.407$ at the optimal~$\chi_{\rm const}$.

%It is hard to see from Fig. \ref{fig:figure2} that the realization $\chi_{1}(F)=\chi_{2}(F)$ cannot occur for $\chi_{1}\neq \chi_{2}$. 

\subsection{Numerical analysis for a noisy blackbox}
\begin{figure}[t]
	\centering
	\includegraphics{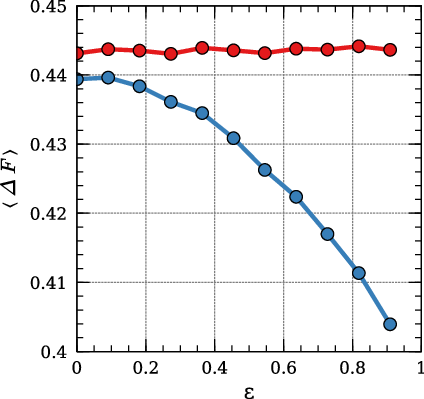}
	\caption{ Upper line: average increase in the similarity for optimal $\chi$ calculated by using~(\ref{chi-def}) without noise in the blackbox. Bottom line: average increase in the similarity for blackbox with error $\varepsilon$ calculated by using~(\ref{chi_opt_1}) and~(\ref{chi_opt_2}). Each point represents averaging over $10^{5}$ random realizations of target states with $b_k$ uniformly distributed in the interval $(-0.5; 0.5)$. The small difference between two lines at~$\varepsilon=0$ shows that our approximation based on~(\ref{chi_opt_1}) and~(\ref{chi_opt_2}) works well.}
	\label{fig:figure4}
\end{figure}

The blackbox operation can involve some noise, i.e. instead of the exact similarity~$F$ we can get a somewhat different $F^{\prime}$ with
\begin{equation}\label{epsilonerror}
	|F-F^{\prime}|<\varepsilon/2,
\end{equation}
where $\varepsilon$ is an error. In such a case the increase of similarity will not be the maximal. 
%In other words, we might erroneously use an imprecise point $(F, \chi)$ from~Fig. \ref{fig:figure2} or corresponding table, and make non-optimal rotation~(\ref{U-rotation}). 
%However, our estimation shows that the average error in~$\chi$ is only~$\approx 0.06$ for $\varepsilon = 0.05$, and $\approx 0.08$ for $\varepsilon = 0.1$ which is much less than the median value of~$\chi$, ~$\chi_{\rm max}/2 = \arctan0.5$.
In Fig.~\ref{fig:figure4} we show the results of simulation of noisy blackbox. We assume that noise induced similarity value~$F'$ has a uniform random distribution around correct value~$F$, and width of this distribution is~$\varepsilon$~(\ref{epsilonerror}). We made~$10^5$ trials for each value of~$\varepsilon$, and from Fig.~\ref{fig:figure4} one can see that our algorithm works also with noisy blackbox.
\begin{comment}
\begin{figure}[t]
	\centering
	\includegraphics{DF-F.eps}
	\caption{Increase in the similarity $\Delta\,F$ after optimal global rotation.}
	\label{fig:figure3}
\end{figure}
\end{comment}

\section{Discussion}
Here we discuss relation to possible experiments. 
To begin with, let us assume that we have two copies of the candidate system. 
In the limit~$J\ll1$ the states of the spin chains are almost disentangled, and in the limit 
$N\gg1$ our blackbox can be realized via projective measurements on the candidate system. 
In this situation the state of the $k$th spin of the candidate (as well as of the target) system is approximately 
pure, i.e. $\ket{\psi_0^{[c,t]}}\approx \ket{\psi_{1}^{[c,t]}}\otimes\ket{\psi_{2}^{[c,t]}}\otimes\dots\ket{\psi_N^{[c,t]}}$, 
and for the candidate system we can write
$$\ket{\psi_{k}^{[c]}}=\cos(\theta_{k}/2)\ket{\psi_{k}^{[t]}} + \sin(\theta_{k}/2)e^{i\phi_{k}}\ket{\psi_{\perp k}^{[t]}},$$ 
with $\braket{\psi_{k}^{[t]}|\psi_{\perp k}^{[t]}} =0 $, and where $\phi_{k}$ is an 
unimportant for our purposes phase. 
In this case $\cos\theta_k = 2 f_k - 1$, where $f_k$ is a commonly defined fidelity~$f_k = |\braket{\psi_{k}^{[t]}|\psi_{k}^{[c]}}|^2$.
We apply projective measurement $\ket{\psi_{k}^{[t]}}\bra{\psi_{k}^{[t]}}$ to 
each spin number~$k$ of the candidate system. The result of each measurement is 
stored in a variable~$m_{k}$: if the result of measurement of the $k$th spin is ``yes'' then $m_{k}=1$ otherwise $m_{k}=0$.
The probability of getting $m_{k}=1$ is $\cos^2(\theta_{k}/2) = (\cos\theta_{k}+1)/2$. 
Thus we have 
$$
F\approx 2\sum_{k=1}^Nm_{k} - N.
$$ 
Due to the probabilistic nature of the blackbox we can expect the dispersion 
(or the error) of $F$ being of the order of $\sqrt{N}\ll N$, where we assume 
that the blackbox does not include the target system physically. This system 
can be ``programmed'' into the blackbox by the input of the $b_{k}$ set. 
The proposed scenario can be assumed as some kind of the ``global quantum state tomography'' which requires only one copy of the large system and which output is the similarity value.
Notice that under the limit $J\ll1$ it is easy to 
compute data $\chi(F)$ (see Fig. \ref{fig:figure2}) for large~$N$, considering spins being independent. 

Such a blackbox for small systems can be realized experimentally by quantum state tomography~\cite{WuByrd} with many copies of 
the target-candidate systems. The reduced density matrices are therefore given,  so is the similarity~$F$. 
As a result, the operations with this blackbox become a limited resource. 

\begin{figure}
	\centering
	\includegraphics[width=80mm]{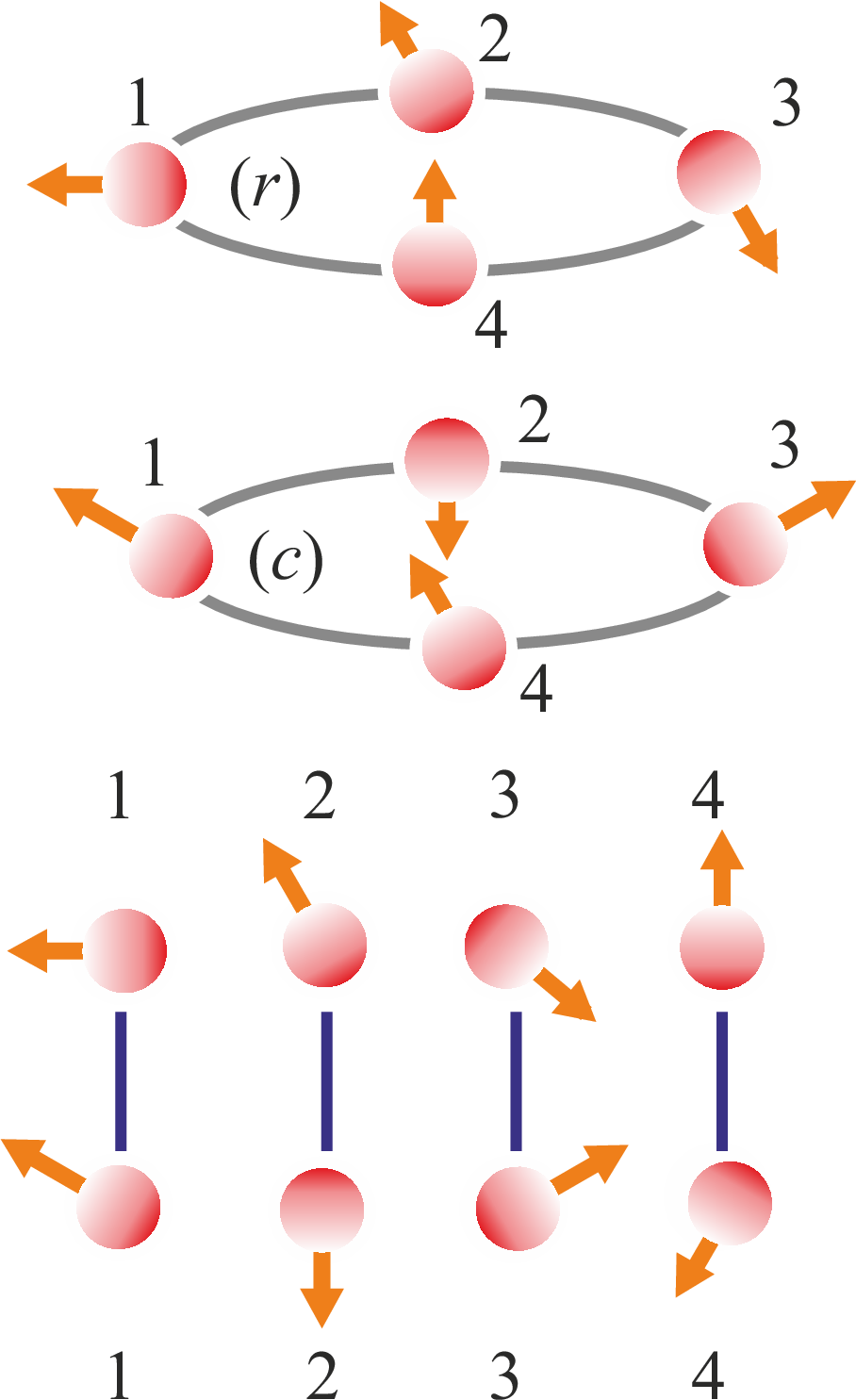}
	\caption{Spin chains transformed into array of dimers.}
	\label{fig:dimers}
\end{figure}

The above discussed problem can be utilized, for example, in the scenario when the corresponding 
spins in two chains (candidate and target) become connected to each other while 
internal interaction $J$ inside each chain is switched off (see Fig. \ref{fig:dimers}), thus forming
ladders of dimers (see, e.g., Ref. \cite{PhysRevLett.113.237203}). If the Hamiltonian of each dimer is ferromagnetic, 
we see that the maximal similarity before the transformation corresponds to the minimal energy 
of the system of dimers.

Finally, we mention that the similarity~(\ref{general_fidelity}) which takes into account some entanglement properties can use, for example, these functions~$f_{\{\bm k\}}$

\begin{equation}\label{general_fidelity_{2}}
	f_{\{\bm k\}} =  1 - \left[\tr\left({\bm\rho}_{\{\bm k\}}^{[t]} \right)^2 - \tr\left({\bm\rho}_{\{\bm k\}}^{[c]}\right)^2\right]^2,
\end{equation}
corresponding to the maximal similarity when entropy (purity) of each 
subset of the candidate system is equal to entropy (purity) of the corresponding subset 
of the target.

\section{Conclusion}

We have studied the manipulation of similarity of quantum states in spin systems by external magnetic
field and demonstrated that even a single step of global rotation can improve similarity related to an unknown state. Importantly, the only information needed from the target system's unknown state is the value of similarity at a single time point when we inquire~(\ref{fidelity_for_chain}) 
which does not provide us information about the structure of the target state. %Lianao: Pavel,  I dont understand this sentence, please reorganize it
Thus, we demonstrated here by a model analysis the possibility of 
accurate mimicking of the initially unknown state by using limited resources. 
Another goal of this research is to formulate the problem of increasing similarity of complex system. 
Given similarity definition using a summation over 
subsystems avoids orthogonality catastrophe which can appear with the commonly defined 
fidelity, and thus it is useful for quantum protocols. Finding the optimal solution of 
the proposed problem without limitation of simplified global control can be of interest for the future
analytical and numerical analysis.

\section*{Acknowledgments}
This work is financially supported through the Grant PGC2018-101355-B-I00 
funded by MCIN/AEI/10.13039/501100011033 and by ERDF ``A way of making Europe'', 
and by the Basque Government through Grants No. IT986-16 and IT1470-22.
P.P.~acknowledges funding by the Free State of Bavaria for the Institute for Topological Insulators.

%\section{Version}

%\label{sec:version}
%This is quantumarticle version v\quantumarticleversion.

\bibliographystyle{iopart-num}
%\bibliography{biblioteka}

\providecommand{\newblock}{}

\end{document}